\documentclass[%
aps,
 amsmath,amssymb,
 reprint,%
]{revtex4-2}

\usepackage{graphicx,epsfig,epsf,float,amsmath,amssymb}
\usepackage{graphicx}% Include figure files
\usepackage{dcolumn}% Align table columns on decimal point
\usepackage{bm}% bold math
\usepackage{hyperref}% add hypertext capabilities
\usepackage{subcaption}
\usepackage{braket}
\usepackage{scrextend}
\usepackage{float}
\usepackage{natbib}

\usepackage{amsfonts,stmaryrd} % Math packages

\usepackage{enumerate} % Custom item numbers for enumerations

\usepackage[ruled]{algorithm2e} % Algorithms

\usepackage[framemethod=tikz]{mdframed} % Allows defining custom boxed/framed environments

\usepackage{listings} % File listings, with syntax highlighting

\usepackage{multirow}
\usepackage{color, colortbl}
\usepackage{makecell}
\usepackage{mathtools}
\usepackage{fontenc}
\usepackage{graphicx}
\usepackage{subcaption}
\usepackage{XCharter} % Use the XCharter fonts

\begin{document}

%%%%%
%
% Please provide the following information
%
%%%%%
\title{Exact solvability of the Gross-Pitaevskii equation for bound states subjected to general potentials}
\author{M. Mirón and E. Sadurn\'i    }
\affiliation{Instituto de Física, Benemérita Universidad Autónoma de Puebla, Apartado Postal J-48, 72570 Puebla, Mexico }
%%%%%
%
% Use as many authors and addresses as required
%
%%%%%

\begin{abstract}
 In this paper we present the analytic solution to the problem of bound states of the Gross-Pitaevskii (GP) equation in 1D and its properties, in the presence of external potentials in the form of finite square wells or attractive Dirac deltas, as well as stable solitons for repulsive defects. We show that the GP equation can be mapped to a first-order non-autonomous dynamical system, whose solutions can sometimes be written in terms of known functions. The formal solutions of this non-conservative system can be written with the help of Glauber-Trotter formulas or a series of ordered exponentials in the coordinate $x$. With this we illustrate how to solve any nonlinear problem based on a construction due to Mello and Kumar for the linear case (layered potentials). For the benefit of the reader, we comment on the difference between the integrability of a quantum system and the solvability of the wave equation.   
\end{abstract}

\maketitle

\section{Introduction}

In this paper we employ a method of quadratures to solve the nonlinear Schrödinger equation in 1D for bound states in a general potential, i.e. solitons with discrete spectra. It is sometimes suggested that only a finite list of potentials allow analytical expressions for stationary waves and energies. This statement is usually motivated by the list of potentials that allow factorization in the sense of Infeld and Hull \cite{infeld}. Here, however, we give a formal solution to any stationary problem with bound states by working out the specific cases of constant piecewise potentials and point-like defects; then the results are generalized to any potential by taking the limit of an infinite number of coalescent regions. This shows that explicit wave functions can always be found, as well as transcendental equations that determine the energy eigenvalues, albeit the need of numerical evaluations for obtaining the spectrum. This method is in compliance with the definition of boundary conditions at infinity.

From the first treatments of the nonlinear Schr\"odinger equation \cite{Dalfovo, Griffin, Frye, Chin, Kraemer, Moerdijk, Bradley,Roberts} it was already evident that nonlinearities could pose additional challenges for the computation of waves and spectra. The impact of potentially new analytical solutions of this problem reaches many areas with diverse applications \cite{Kerr1, Kerr2, London, Higgs, Goldstone, Altarelli}. Although it is known that the Jacobi Elliptic functions solve the nonlinear equation with piecewise constant external potentials \cite{Miron, Gradshteyn}, the method of wave function matching across boundaries is applicable only when full multi-parametric solutions are given explicitly; in this paper, we show that this can be done by our quadrature method. In the linear case, there is a counted list of potentials with closed solutions for energies and wave functions \cite{Ginocchio, Philip, Eckart, Poschl, Natanzon}, as well as their super symmetric extensions \cite{Cooper}, but so far nothing has been reported for the Gross-Pitaevskii (GP) equation (with the exception of the so-called Thomas-Fermi limit, which is equivalent to a strong nonlinearity in the GP case). 

In order to put our contribution in context, it is important to mention that the most general form of stationary solutions for the linear case is simply given by the $2\times2$ scattering matrix and its pole structure, even for bound states with vanishing conditions at infinity and purely imaginary poles in the $k$ plane. Formally, one can write the solutions of second-order differential equations by mapping the system to a two-dimensional dynamical problem \cite{Mostafazadeh1,Mostafazadeh2} of a lesser degree, i.e. a vector first-order equation. The idea of interpreting the coordinate $x$ as a quasi-time was recently given in \cite{Miron}. For the case of a general piecewise potential, gluing all pieces together by boundary matching can be done with square potentials, as well as a series of Dirac deltas, whichever is convenient. For the linear problem, the corresponding energy-dependent Green's functions and their connection with the scattering matrix are well known \cite{Grosche, Castro-Alatorre, Moshinsky}, but in the nonlinear case there are no such constructions; the Green's function is not available. It is interesting to note that Mello and Kummar indicated a layered construction for the general scattering problem \cite{Mello} that, for convenience, can be reduced to a differential equation for the \textit{scattering matrix} and not for the \textit{wave function}. Now, in a similar construction, we generalize the treatment for the nonlinear case and the wave function.

Perhaps the most important motivation of these studies, in recent times, belongs to the quantum mechanical realization of the GP equation in Bose-Einstein condensation (BEC) and the existence of multiple nonlinear bound states in a potential well or lattice trap \cite{Bloch}. These bound solitons, mentioned in \cite{Miron, Carr,Rajaraman}, undergo multilevel transitions that may be employed in the construction of qubits, qutrits and the like. Various numerical methods to attack the problem are given in \cite{Bao1, Symes, Bao2, LeVeque}, also in connection with spinorial BEC generalizations and the application of magnetic fields.

Structure of this paper: In Section 2 we briefly review the quadrature method. In Section 3 we solve the bound state problem for a delta defect and a square well potential using Jacobi Elliptic functions. The energies are obtained by solving a transcendental equation with the graphic method. In Section 4 we generalize the method to arbitrary potentials in the continuous limit of layered defects; we do this for the linear as well as the nonlinear problem, arriving at a dynamical set of equations in the spatial coordinate. In Section 5 we discuss the general solutions in the context of classical and quantum-mechanical integrability. Conclusions are drawn in Section 6.

\section{Gross-Pitaevskii equation}
\begin{figure}[H]
\centering
 \includegraphics[width=0.65\linewidth]
    {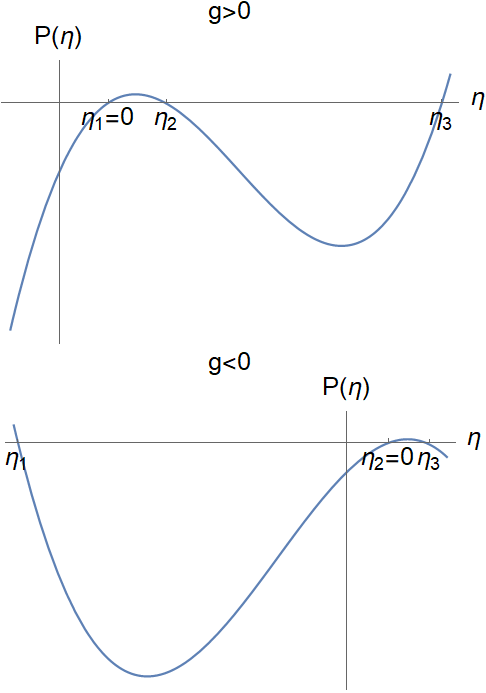}
 \caption{Roots of the polynomials that define the elliptic Jacobi functions.}
 \label{FIG10}
\end{figure}

The dynamical behavior of a BEC trapped in 1D, in the presence of an external potential $V_{\mathrm{ext}}(x)$, is given by the time-dependent GP equation. In stationary form, we work with the differential equation
\begin{equation}
   \left\lbrace -\frac{\hbar^{2}}{2m} \frac{d^{2}}{dx^{2}}+g|\phi(x)|^{2} +V_{\mathrm{ext}}(x)\right\rbrace \phi(x)=E \phi(x).
\label{PE}
\end{equation}
This is deduced from a mean field theory considering only contact interactions, the particles are in the same state, and the state function of a single particle is sufficient to describe the complete bosonic system. Here, $m$ is the particle's mass, $\hbar$ is the reduced Planck constant, $E$ is the energy of the stationary state, and $g$ is related to the scattering length $a$ of $s-$state through $g=(4 \pi \hbar^{2} a)/m$ \cite{Dalfovo}. The map to a dynamical system with quasi-time $\tau = \sqrt{2m} x / \hbar$ (proportional to $x$) is defined by a complex coordinate $X(\tau) = \phi(x)$ and its velocity $\dot X(\tau) = d \phi(x) / dx$. A first-order differential equation emerges, involving a complex vector with components $(\dot X, X)$:

\begin{align}
 \frac{d}{d \tau} \left( \begin{array}{c}
      \dot X  \\
      X 
 \end{array} \right) = \left( \begin{array}{cc}
      0 & V-E-\Phi[X] \\
      1  & 0 
 \end{array} \right)  \left( \begin{array}{c}
       \dot X  \\
     X 
 \end{array} \right), 
\end{align}
with $\Phi[X,\tau] = (2m/\hbar^2)[ (E-V(\tau)) |X|^2/2 -g |X|^4/4]$. When the potential $V$ is constant, the following quasi-energy functional is conserved (in $\tau$ or $x$):

\begin{align}
U = |\dot X|^2 /2 + \Phi[X].    
\end{align}
It is important to note that the complex character of $X(\tau)$ makes the system two-dimensional, and in this kind of mapping, the polar coordinates $(r,\varphi)$ in the complex plane represent the density $r^2 = |\phi|^2$ and the phase $\varphi = \rm{arg}(\phi)$. In general, $\varphi$ does not vanish, and its variation can be identified with the conserved probability current $J = r^2 d \varphi / dx, \; d J / dx = 0 $, which is in full parallel with the conserved angular momentum of a particle in a planar space under the influence of an isotropic potential $ \Phi[r]=(2m/\hbar^2)(E |\phi|^2 - g |\phi|^4) $. Using the separability of the problem, the radial coordinate \textit{feels} the action of an effective potential with a centrifugal barrier $ \Phi_{\rm eff}[r]= \Phi[r] + (2m/\hbar^2)(J^2/r^2)$. As we have shown in previous work, scattering solutions necessitate a non-vanishing $\varphi$ and its inherent phase shift, but since bound states are the main focus of the present discussion, we can impose $\varphi = 0$. The quadrature method consists in evaluating the integral for the period or lapse $\tau$ and solve for $r$ alone, where $E$ and $r_0$ are the only parameters:
\begin{align}
\tau = \int_{r_0} \frac{dr}{\sqrt{2m(U-\Phi[r])}}.
\label{quad}
\end{align}
This greatly simplifies the equations above, as well as the effective potential; we have $ \Phi_{\rm eff}[r]= \Phi[r]$ and $X$ real. Under a change of variables $\eta = r^2$, this quadrature (\ref{quad}) is transformed into the integral representation of the Jacobi Elliptic function, where the limits of integration are compatible with the turning points of the potential $\Phi$, i.e., where the expression inside the radical is a third-order polynomial $P(\eta)$ with real roots and positive values. Depending on the sign of $g$, the polynomial $P(\eta)$ has a specific ordering of its roots $\eta_i, i=1,2,3$. This is shown in figure \ref{FIG10}. In what follows, we illustrate the method of solution for particles bound by simple potentials using boundary matching conditions.
\section{Nonlinear bound states in delta and box potentials}
Two simple models are solved in order to illustrate the boundary matching method in nonlinear systems; from these two examples, we may construct more complicated cases.
\subsection{Dirac delta}
We solve the GP equation for bound states in 1D. As a result, we shall find that, contrary to the linear case, positive Dirac deltas can support bound states up to a threshold negative coupling $g$. Similarly, a negative delta may lose its binding capabilities if a positive nonlinear coupling $g$ is sufficiently strong. We start with the equation
\begin{equation}
\left\lbrace -\frac{d^{2}}{d \tau^{2}}+V_{\mathrm{ext}}(\tau)+\gamma \psi^{2}(\tau) \right\rbrace \psi(\tau) = E \psi(\tau),
\end{equation}
where a convenient rescaling is used $\tau=\sqrt{2m} x/\hbar$, $\bar{\alpha}=\sqrt{2m} \alpha/\hbar$, $\gamma=\sqrt{2m} g/\hbar$, $V_{\mathrm{ext}}(\tau)=\bar{\alpha} \delta(\tau)$, $ \psi= (2m)^{1/4} \phi / \sqrt{\hbar}$. We divide the problem into two regions: 1 denotes $\tau<0$ and 2 corresponds to $\tau>0$. Then the matching conditions are given by $\phi_{2}(0)=\phi_{1}(0), \; \dot{\phi}_{2}(0)=\dot{\phi}_{1}(0)+\bar{\alpha} \phi_{1}(0)$.
The following two cases can be distinguished: generalized bright solitons and logarithmic quadratures, which are shown below.
\subsubsection{Bright solitons} 
For bright solitons, $E<0$ and $\gamma<0$, there are bound states given by 
\begin{itemize}
\item[1.] \textit{Repulsive defect}
\begin{equation}
\phi_{\bar{\alpha}>0}(\tau)=\left\{ \begin{array}{lcc} \sqrt{\frac{2 E}{\gamma}} \mathrm{sech} \left(\sqrt{|E|} (\tau+\bar{\tau}) \right) & \mathrm{if} & \tau<0 \\ 
\\ \sqrt{\frac{2 E}{\gamma}} \mathrm{sech} \left( \sqrt{|E|} (\tau-\bar{\tau}) \right)  & \mathrm{if} & \tau>0 \end{array} \right.
\end{equation}
with $\bar{\alpha}=2 \sqrt{|E|} \tanh{(\sqrt{|E|} \bar{\tau})}$ and $0 \leq \bar{\alpha} \leq 2 \sqrt{|E|}$.

\item[2.] \textit{Attractive defect}
\begin{equation}
\phi_{\bar{\alpha}<0}(\tau)=\left\{ \begin{array}{lcc} \sqrt{\frac{2 E}{\gamma}} \mathrm{sech} \left(\sqrt{|E|} (\tau-\bar{\tau}) \right) & \mathrm{if} & \tau<0 \\ 
\\ \sqrt{\frac{2 E}{\gamma}} \mathrm{sech} \left( \sqrt{|E|} (\tau+\bar{\tau}) \right)  & \mathrm{if} & \tau>0 \end{array} \right.
\end{equation}
with $\bar{\alpha}=-2 \sqrt{|E|} \tanh{(\sqrt{|E|} \bar{\tau})}$ and $- \infty \leq \bar{\alpha} < 0$.
\end{itemize}
In both cases
\begin{equation}
E=- \frac{\bar{\alpha}^{2}}{4}+\frac{\gamma}{2} \phi^{2}(0). 
\end{equation}

\begin{figure}[H]
 \includegraphics[width=\linewidth]
    {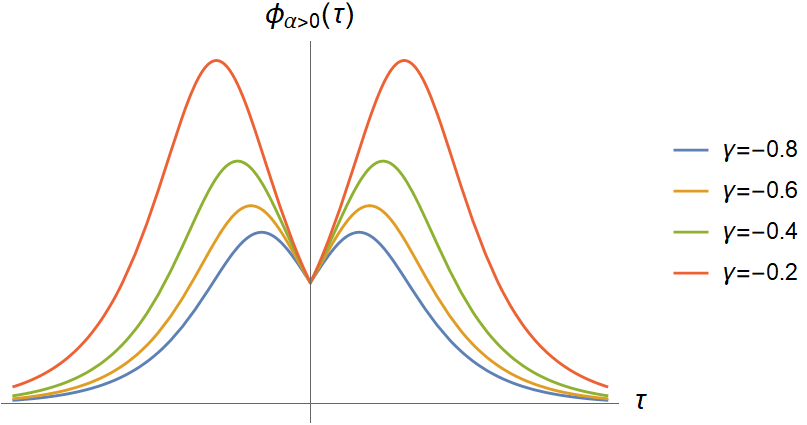}
 \caption{Bright solitons in repulsive barrier. $\phi_{\alpha>0}(0)$ and $\alpha$ are fixed. When $\gamma \rightarrow 0$, $\phi_{\bar{\alpha}>0}(-\bar{\tau})=\phi_{\bar{\alpha}>0}(\bar{\tau}) \rightarrow \infty$ and $\bar{\tau} \rightarrow \infty$.}
 \label{familia}
\end{figure}

\begin{figure}[H]
\centering
 \includegraphics[width=0.8\linewidth]{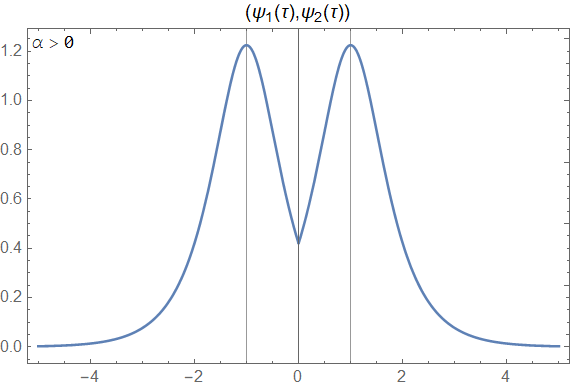}
 \caption{Bright solitons. Repulsive defect and negative coupling. The bound state is repelled but not destroyed by the point-like potential. }
 \label{FIG1}
\end{figure}

\begin{figure}[H]
\centering 
 \includegraphics[width=0.8\linewidth]{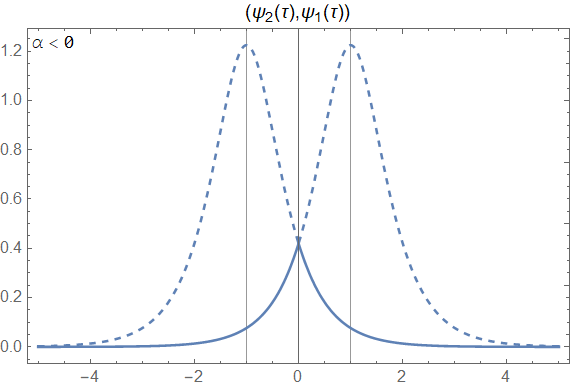}
 \caption{Bright solitons. Attractive defect and negative coupling. Both contributions make the binding mechanism stronger.}
 \label{FIG2}
\end{figure}

The results are displayed in figs. \ref{FIG1}  and \ref{FIG2}, where the effect of the delta potential changes the shape of the soliton according to its sign. In the repulsive case, we observe that in the limit when $\gamma \rightarrow 0$, the energy remains negative $E \rightarrow -\alpha^{2}/4$, if the amplitude at the defect is kept constant, for which $\phi_{\bar{\alpha}>0}(-\bar{\tau})=\phi_{\bar{\alpha}>0}(\bar{\tau}) \rightarrow \infty$ and $\bar{\tau} \rightarrow \infty$, i.e., the amplitude and the two maxima of the soliton increase, as we can see in fig. \ref{familia}.

\subsubsection{Solution associated to a logarithmic quadrature} 
\begin{figure}[H]
\begin{center}
 \includegraphics[width=0.7\linewidth]{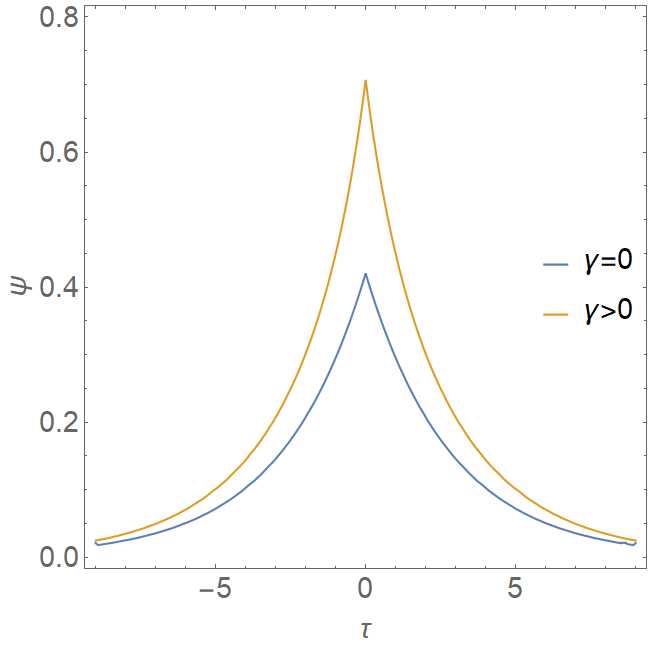}
 \caption{Wave functions computed with a logarithmic quadrature, valid for positive nonlinear coupling. When the densities are properly normalized, the localization length of $\gamma=0$ and $\gamma>0$ can be compared, and the loosely bound state can be identified by its larger width.}
 \label{FIG3}
\end{center} 
\end{figure}

Here we have  $\bar{\alpha}<0$, $E<0$ and $\gamma>0$, representing bound states. In this case, the equation for quasi-energy conservation becomes
\begin{eqnarray}
    U_{1}=\frac{1}{2} \dot{\phi_{1}}^{2}(\tau)+\Phi_{\mathrm{eff}}(\phi_{1}), \; \tau<0; \\
U_{2}=\frac{1}{2} \dot{\phi_{2}}^{2}(\tau)+\Phi_{\mathrm{eff}}(\phi_{2}), \; \tau>0; \\ 
\Phi_{\mathrm{eff}}(\phi_{i})=\frac{1}{2} E \phi^{2}_{i}(\tau)-\frac{\gamma}{4} \phi^{4}_{i}(\tau).
\end{eqnarray}
The only admissible solution is obtained with $U_{1}=0=U_{2}$, $\phi_{1}(0)=\phi_{2}(0)=\phi_{0}$, $\dot{\phi}_{1}(0)=(\bar{\alpha} \phi_{0})/2=-\dot{\phi}_{2}(0)$, $\phi_{0}=\sqrt{(\bar{\alpha}^{2}-4|E|)/2 \gamma}$, i.e., 
\begin{align}
-\sqrt{|E|}|\tau|= \log \left\lbrace \frac{(1+2\sqrt{|E|})\psi(\tau)}{\sqrt{1-4|E|} \left(\sqrt{\frac{2|E|}{\gamma}} + \sqrt{\psi^{2}(\tau)+\frac{2|E|}{\gamma}}\right)} \right\rbrace, 
\label{logquad}
\end{align}
under the conditions $\bar{\alpha} \leq -\sqrt{2 \gamma}\phi_{0}=\alpha_{\mathrm{crit}}$, $E=\gamma \phi^{2}(0)/2-\bar{\alpha}^{2}/4 \leq 0$. From (\ref{logquad}) it is possible to recover the explicit wave function in terms of exponentials or hyperbolic functions, by trivially solving for $\psi(\tau)$. We show the results in fig. \ref{FIG3}. The intensity parameter $\gamma$ of the nonlinearity modifies the width of the soliton, hence the localization of the wave, for a fixed $\bar \alpha$.

\noindent

It is important to note that bound states exist if $\bar{\alpha}$ satisfies the inequalities above. This is in sheer contrast with the linear case, where any attractive potential in 1D is able to bind a particle (in 2D or higher dimensions, an attractive potential needs a critical depth to produce at least one bound state). The nonlinear case demands sufficient attraction to surpass the positive contribution $\gamma \phi^{2}_{0}$. 

\subsection{Square box}
\begin{figure}[H]
 \includegraphics[width=\linewidth]
    {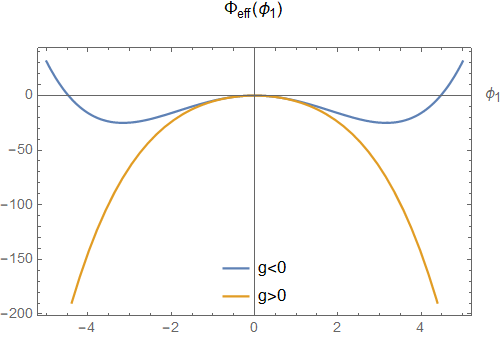}
 \caption{Effective potential $\Phi_{\rm eff}$ outside of the well $V_{\mathrm{ext}}$.}
 \label{FIG4}
\end{figure}

\begin{figure}[H]
 \includegraphics[width=\linewidth]
    {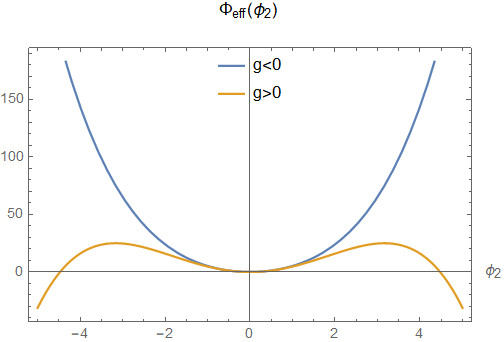}
 \caption{Effective potential $\Phi_{\rm eff}$ inside of the well $V_{\rm ext}$.}
 \label{FIG5}
\end{figure}

The boundary matching method divides the problem into three regions; 1 and 3 correspond to the exterior of the trap and 2 corresponds to the interior. As in the linear case, the continuity of the wave function and its derivative provide a transcendental equation for the allowed energy eigenvalues. For the nonlinear problem, this equation is a generalization of the familiar relation $k \tan k = \rm{const}$ in the linear case, and shall be derived in terms of Jacobi amplitudes below. The graphic method of solution will be presented as well.
The GP equation with a square box potential is given by
\begin{align}
& \left\lbrace -\frac{d^{2}}{d \tau^{2}}+g\phi^{2}(\tau)+V_{\mathrm{ext}}(\tau) \right\rbrace \phi(\tau)=E \phi(\tau), \; \; \tau=\frac{\sqrt{2m}}{\hbar} x, \\
& V_{\mathrm{ext}}(\tau)=\left\{ \begin{array}{lcc}  -V_{0} & \mathrm{if} & |\tau| \leq \frac{\sqrt{2m}}{\hbar} x_{0}=\tau_{0} \\ 
\\ 0  & \mathrm{if} &  |\tau| > \frac{\sqrt{2m}}{\hbar} x_{0}=\tau_{0} \end{array} \right., \; V_{0}>0.
\end{align}
The quasi energy in the three regions and the effective potential are
\begin{eqnarray}
U_{i}=\frac{1}{2} \dot{\phi}^{2}_{i}(\tau)+\Phi_{\mathrm{eff}}[\phi_{i}(\tau)]=\mathrm{const.}, \\
\Phi_{\mathrm{eff}}[\phi_{i}(\tau)]=-\frac{g}{4} \phi^{4}_{i}(\tau)+\frac{1}{2}[E-V(\tau)]\phi^{2}_{i}(\tau),
\end{eqnarray}
where $i=$ 1, 2 and 3. We define $\phi_{1}(\tau)$ and $U_{1}$ in $\tau<-\tau_{0}$, $\phi_{2}(\tau)$ and $U_{2}$ in the region $-\tau_{0} \leq \tau \leq \tau_{0}$, and $\phi_{3}(\tau)$ and $U_{3}$ in $\tau_{0}<\tau$. The plots of the effective potentials are similar to the well-known \textit{Mexican-hat} shape and are shown in figs. \ref{FIG4} and \ref{FIG5}.
The boundary matching relations are, in this case, strictly continuous:
$\phi_{1}(-\tau_{0})=\phi_{2}(-\tau_{0})$, $\dot{\phi}_{1}(-\tau_{0})=\dot{\phi}_{2}(-\tau_{0})$, $ \phi_{2}(\tau_{0})=\phi_{3}(\tau_{0})$ and $\dot{\phi}_{2}(\tau_{0})=\dot{\phi}_{3}(\tau_{0})$. Since the external potential is symmetric, there are two families of solutions given by symmetric functions, which fulfill $\dot{\phi}_{2}(0)=0$ and antisymmetric functions, with the property $\phi_{2}(0)=0$. Then the two cases $g>0$ and $g<0$ are treated separately as indicated in the following.

\subsubsection{Positive nonlinear coupling $g$}
Using the elliptic integral and consistency with matching relations at $\pm \tau_0$ leads, for $\tau \leq -\tau_{0}$, to
\begin{eqnarray}
&  -\sqrt{|E|}(\tau_{0}+\tau)= \nonumber \\ 
& \log \left\lbrace \frac{ \left[ \sqrt{\frac{2|E|}{g}}+\sqrt{\phi^{2}_{1}(\tau)+\frac{2|E|}{g}} \right]}{\left[ \sqrt{\frac{2|E|}{g}}+\sqrt{\phi^{2}_{1}(-\tau_{0})+\frac{2|E|}{g}} \right]} \frac{\phi_{1}(-\tau_{0})}{\phi_{1}(\tau)}\right\rbrace,
\end{eqnarray}
while $-\tau_{0} \leq \tau \leq 0$ yields
\begin{align}
&  \phi_{2}(\tau)=\sqrt{\eta_{2}} \cos \left\lbrace \alpha_{1}(\tau) \right\rbrace, \nonumber \\ 
& \alpha_{1}(\tau)=\mathrm{Am} \left[ \sqrt{\frac{g \eta_{3}}{2}} (\tau+\tau_{0}) +\mathrm{F} \left( \arcsin \left\lbrace \frac{\phi_{1}(-\tau_{0})}{\sqrt{\eta_{2}}} \right\rbrace, \right. \right. \nonumber \\ 
& \left. \left. \sqrt{\frac{\eta_{2}}{\eta_{3}}} \right), \sqrt{\frac{\eta_{2}}{\eta_{3}}} \right]-\frac{\pi}{2}.
\end{align}
Here, the roots $\eta$ associated with the turning points of $\Phi$ are given by
\begin{align}
& \eta_{1}=0, \; \eta_{2}=\frac{E+V_{0}}{g}-\sqrt{\left(\frac{E+V_{0}}{g} \right)^{2}-\frac{2 V_{0}}{g} \phi^{2}_{1}(-\tau_{0})}, \nonumber \\ 
& \eta_{3}=\frac{E+V_{0}}{g}+\sqrt{\left(\frac{E+V_{0}}{g} \right)^{2}-\frac{2 V_{0}}{g} \phi^{2}_{1}(-\tau_{0})}.
\end{align}
For the symmetric case, we make use of the cosine function of $\alpha$ in the interior, giving rise to the energy quantization condition $\alpha_{1}(0)=n \pi$, $n=0,1,2,...$, and for the antisymmetric case $\alpha_{1}(0)=(n+1/2) \pi$, $n=0,1,2,...$. These conditions are derived here for the first time, and they lead to the graphic method displayed in figures \ref{FIG6} and \ref{FIG7}, where $\alpha$ is plotted against $E$, and its intersection with the quantized values yields the discrete energies sought for bound states.

\subsubsection{Negative nonlinear coupling $g$}  
\begin{figure}[H]
 \includegraphics[width=\linewidth]
    {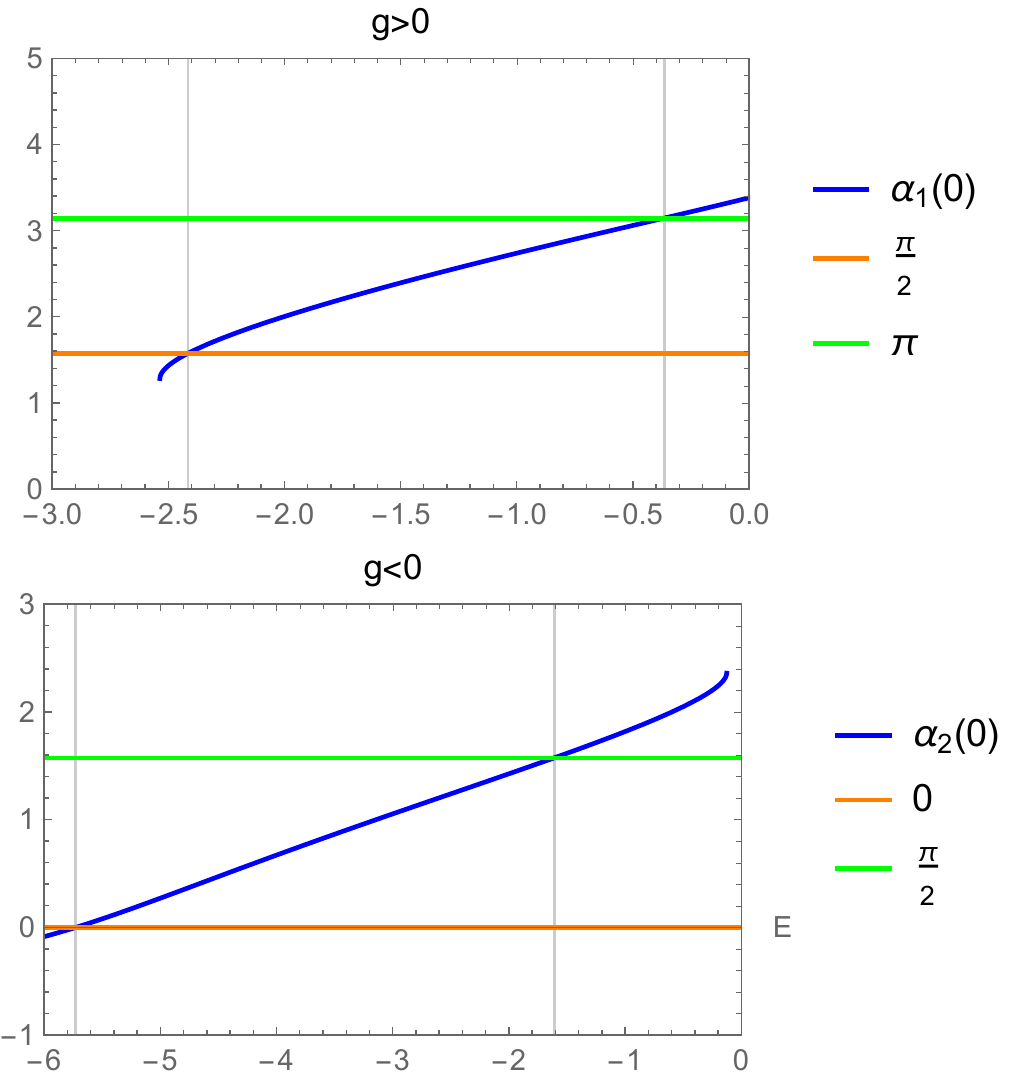}
 \caption{Energy curves in the graphic method of solution. Upper panel, $g>0$: $V_{0}=6$, $g=4$, $\phi_{0}=0.5$, $\tau_{0}=1$. Lower panel, $g<0$: $V_{0}=6$, $g=-1$, $\phi_{0}=0.5$, $\tau_{0}=1$.}
 \label{FIG6}
\end{figure}

\begin{figure}[H]
 \includegraphics[width=0.9\linewidth]
    {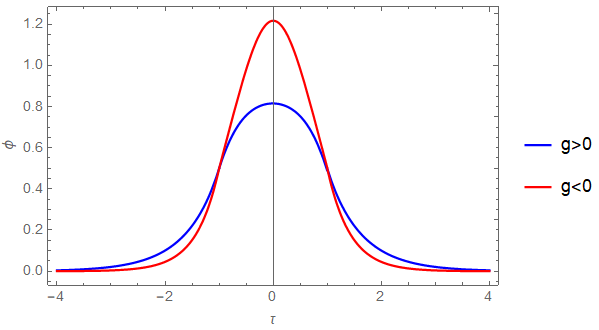}
 \caption{Ground state for both coupling cases. The curvatures display qualitatively different features.}
 \label{FIG7}
\end{figure}

\begin{figure}[H]
 \includegraphics[width=0.9\linewidth]
    {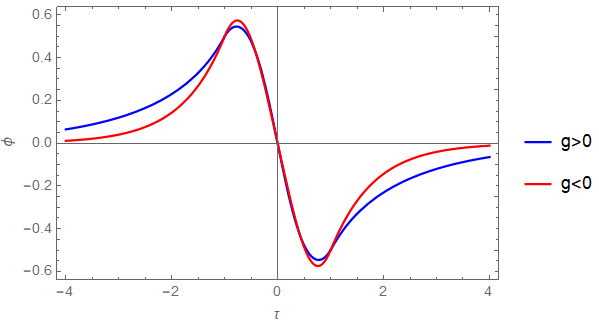}
 \caption{First excited state for both cases, when values of $g$ and $V$ allow its existence.}
 \label{FIG8}
\end{figure}

We proceed as in the previous case, but with different locations for the roots of $P(\eta)$. For $\tau \leq -\tau_{0}$, we have:
\begin{eqnarray}
 \phi_{1}(\tau)=\sqrt{\frac{2E}{g}} \mathrm{sech} \left[\sqrt{|E|}(\tau_{0}+\tau) \right. \nonumber \\ 
\left. +\mathrm{arcsech} \left( \sqrt{\frac{g}{2E}} \phi_{1}(-\tau_{0}) \right) \right],
\end{eqnarray}
and in the region $-\tau_{0} \leq \tau \leq 0$,
\begin{align}
&  \phi_{2}(\tau) = \sqrt{\eta_{3}} \cos \left\lbrace \alpha_{2} (\tau)\right\rbrace, \nonumber \\ 
& \alpha_{2} (\tau)=\mathrm{Am} \left[ \sqrt{\frac{|g| (\eta_{3}-\eta_{1})}{2}} (\tau+\tau_{0}) \right.  \nonumber \\ 
& \left. - \mathrm{F} \left( \arcsin \left\lbrace \sqrt{\frac{\eta_{3}-\phi^{2}_{1}(-\tau_{0})}{\eta_{3}}} \right\rbrace, \sqrt{\frac{\eta_{3}}{\eta_{3}-\eta_{1}}} \right), \right. \nonumber \\
& \left. \sqrt{\frac{\eta_{3}}{\eta_{3}-\eta_{1}}}  \right],
\end{align}
with the following explicit forms of the turning points:
\begin{eqnarray}
& \eta_{1}=\frac{E+V_{0}}{g}-\sqrt{\left(\frac{E+V_{0}}{g} \right)^{2}+\frac{2 V_{0}}{|g|} \phi^{2}_{1}(-\tau_{0})}, \; \eta_{2}=0, \nonumber \\
& \eta_{3}=\frac{E+V_{0}}{g}+\sqrt{\left(\frac{E+V_{0}}{g} \right)^{2}+\frac{2 V_{0}}{|g|} \phi^{2}_{1}(-\tau_{0})}.
\end{eqnarray}
For the symmetric case, we now have the following quantization condition $\alpha_{2}(0)=n \pi$, $n=0,1,2,...$, while antisymmetric waves require the modified relation $\alpha_{2}(0)= (n+1/2) \pi$, $n=0,1,2,...$. We note here that the symmetric quantum number starts at $\alpha=0$, but this is consistent with the previous case $g>0$ in the limit $g \rightarrow 0$, i.e., the energy curves for the ground state are continuous. Note, however, that fixing the value of the wave functions at $\pm\tau_0$ forces all solutions to depend on this specific amplitude (contrary to the linear case, where overall scales of wave functions are irrelevant) and they are not immune to normalization. Although it is possible to use the total number of particles $\int dx |\phi|^2 = N$ as a parameter, this lengthy expression shall be avoided here for simplicity. As a consequence, the energy levels \textit{are} continuous functions of $g$, but their slope exhibits a \textit{kink} at $g=0$, as can be shown explicitly by differentiating the quadrature with respect to $g$ and employing the chain rule. We provide further comment on this below. The resulting wave functions are plotted in figs. \ref{FIG6}, \ref{FIG7}  and \ref{FIG8} for the ground state and first excited state. The curvature changes significantly as a function of $g$, as well as the width of the distributions (the binding capabilities of the potential depend on the sign of $g$ and can be characterized by a typical localization length conveniently defined by the second moment of the distribution).  

\begin{figure}[H]
 \includegraphics[width=\linewidth]
    {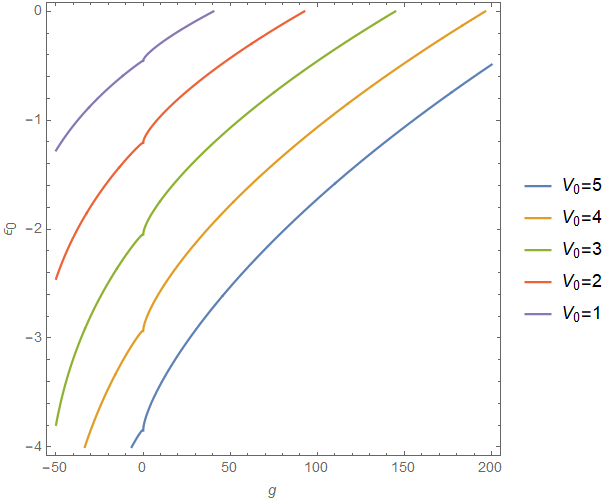}
 \caption{Ground state energy vs coupling for different depths.}
 \label{FIG9}
\end{figure}

In fig. \ref{FIG9} we show the evolution of energy levels with the nonlinear coupling. As expected, there is always a critical value of $g>0$ for which $V$ cannot bind a particle, as the energy curves reach $E=0$ at the top of the plot. We can also observe the kink in the curve $E$ vs $g$ at $g=0$. The curves are continuous, but their derivative has a jump in the transition between attractive and repulsive self-interactions. This can be explained easily by means of the quadrature, since for symmetric and antisymmetric bound states, the functions $\alpha_{1}(0)$ and $\alpha_{2}(0)$ are parameterized in different ways according to the sign of $g$; meanwhile $g$ is contained in both arguments of the elliptic functions $F(\;,\;)$ and $\mathrm{Am}(\;,\;)$ and we must keep track of such a dependence when computing the derivative with respect to $g$ at $g=0$. When $g \rightarrow 0$, we see that $\alpha_{1}(0)=\alpha_{2}(0)$ but $d \alpha_{1}(0)/dg  \neq d\alpha_{2}(0)/dg $.

\section{Formal solution of the GP equation with general potential}
We now address the nonlinear problem in its most general form. Our phase-space differential equations are, again, non-autonomous and their integration can be given through iterative series. It is important to analyze the existence of critical points of such dynamical systems, since these are no longer trivial. Furthermore, it must be recognized that the integral form of the equation is not useful \textit{per se}, as it does not provide a true solution for the wave function before iteration. In this sense it is more useful to use a composition law of ordered exponentials that can be fed from the analytical solutions for the case $g=0$, giving rise to expressions of the Glauber-Trotter type \cite{Trotter} for products of operators. This shall be demonstrated below.

We show that, indeed, all nonlinear 1D problems allow analytical solutions for waves and for the expressions that determine their eigenvalues, but they are merely formal expressions. The existence of solitons or bound spectra represents a greater degree of difficulty in the general case with arbitrary $V$. Our novel method is reduced to the constructive form of the scattering matrix exposed by Mello et al. \cite{Mello}; in fact, it is enough to analyze the transmission and reflection in the plane of complex energies to obtain the solutions of any linear problem according to its boundary conditions.  However, we go one step further by avoiding the use of the scattering matrix (non-existent for GP) and writing the most general possible wave with parametric dependence on the energy.

In general, we consider the external potential as a continuous limit of a Dirac comb
\begin{equation}
V_{\mathrm{ext}} (x)=\sum^{N}_{i=0} v_{i} \delta (x-x_{i}),
\end{equation}
with the definitions $v_{i}=V_{i} \Delta x =V_{\mathrm{eff}}(x_{i}) \Delta x$, $x_{i}=i \Delta x+x_{0}$, and $\Delta x=L/N$; $L$ is the range or support of the external potential. By rescaling eq. (\ref{PE}), for bound states or solitons $\phi$, we obtain
\begin{equation}
    \left\lbrace -\frac{d^{2}}{d \tau^{2}} +g \phi^{2}(\tau)+\sum^{N}_{i=0} v'_{i} \delta(\tau-\tau_{i}) \right\rbrace \phi(\tau)=E \phi(\tau),
\label{GE}
\end{equation}
with $\tau=l x$, $l=\sqrt{2m}/\hbar$, $v'_{i}=\bar{V}_{i} \Delta \tau$, and since $\Delta \tau = l \, \Delta x$, the potential is not rescaled, i.e., $\bar{V}_{i}=V_{i}=V_{\mathrm{eff}}(x_{i})$.
The continuity of the wave function and the discontinuity in its derivative give us the relations
\begin{eqnarray}
\phi_{i+1}(\tau_{i})=\phi_{i}(\tau_{i}), \\ 
\dot{\phi}_{i+1}(\tau_{i})=\dot{\phi}_{i}(\tau_{i})+v'_{i} \phi_{i}(\tau_{i}), \\
\phi_{i}=\phi_{i}(\tau), \; \tau_{i-1} \leq \tau \leq \tau_{i}, \; i=1,...,N.
\end{eqnarray}
In the continuous limit, we recover any desired potential $V$ whose integral is well-defined. We have
\begin{eqnarray}
& \lim_{\scriptsize N \rightarrow \infty,\, \scriptsize \Delta \tau \rightarrow 0}   \sum^{N}_{i=1}  \Delta \tau V_{i} \delta(\tau-\tau_{i}) = \nonumber \\ & \int d\tau' V(\tau') \delta(\tau-\tau')=V(\tau). 
\end{eqnarray}
Our approach is then well justified.
\subsection{Linear case}
We consider first $g=0$. In the interstitial region, between successive deltas, the solution is given by $\phi_{i}(\tau)=A_{i} \sin (\sqrt{E}\tau-\alpha_{i})$, with $\alpha_{i}$ and $A_{i}$ real functions. The boundary matching conditions are
\begin{equation}
A_{i+1} \sin( \sqrt{E} \tau_{i}-\alpha_{i+1})=A_{i} \sin(\sqrt{E} \tau_{i}-\alpha_{i}), 
\label{dos}
\end{equation}
\begin{eqnarray}
   & \sqrt{E} A_{i+1} \cos(\sqrt{E} \tau_{i}-\alpha_{i+1})=\sqrt{E} A_{i} \cos(\sqrt{E} \tau_{i}-\alpha_{i}) \nonumber \\ 
&+V_{i} \Delta \tau A_{i} \sin(\sqrt{E} \tau_{i}-\alpha_{i}).
\label{uno}
\end{eqnarray}
In the continuous limit we have $A_{i} \rightarrow A(\tau)$, $\alpha_{i} \rightarrow  \alpha(\tau)$, $A_{i+1} \rightarrow A(\tau + \Delta \tau)$, and $\alpha_{i+1} \rightarrow  \alpha(\tau+ \Delta \tau)$, such that
{\footnotesize \begin{eqnarray}  
A(\tau + \Delta \tau) \sin(\sqrt{E} \tau-\alpha(\tau+\Delta \tau))=A(\tau) \sin(\sqrt{E} \tau-\alpha(\tau)), \\
\sqrt{E} A(\tau + \Delta \tau) \cos(\sqrt{E} \tau-\alpha(\tau+ \Delta \tau))=\sqrt{E} A(\tau) \times \nonumber \\ 
\times \cos(\sqrt{E} \tau-\alpha(\tau))
 +V(\tau) \Delta \tau A(\tau) \sin(\sqrt{E} \tau-\alpha(\tau))
\end{eqnarray}}
A two-component vector can accommodate the two rows above, with the definition 
\begin{eqnarray}
\mathbf{r} (\tau)=\begin{pmatrix}
\sqrt{E} A(\tau) \cos{(\sqrt{E} \tau-\alpha(\tau))}\\
A(\tau) \sin{(\sqrt{E} \tau-\alpha(\tau))}
\end{pmatrix},
\label{EQ33}
\end{eqnarray}
where the first component corresponds to (\ref{uno}) and the second to (\ref{dos}). We take the limit $\Delta \tau \rightarrow 0$ and with a Taylor   series approximation to first order, we obtain the differential equation
\begin{eqnarray}
\frac{d \mathbf{r}(\tau)}{d \tau}=\hat{M}(\tau) \mathbf{r}(\tau),
\label{EQ34}
\end{eqnarray}
with a matrix operator containing the potential
\begin{eqnarray}
\hat{M}(\tau)=\begin{pmatrix}
0 & V(\tau)-E\\
1 & 0
\end{pmatrix}.
\end{eqnarray}
The general solution of this linear nonautonomous system is given by a series of ordered integrals. We must bear in mind that $\tau$ is a quasi time that stands for position, so the general solution is a position-ordered exponential in the form
%\end{multicols}
%\begin{multicols}
\begin{widetext}
\begin{equation}
\mathbf{r} (\tau) = \hat R \, \mathbf{r}_{0}, \quad \hat R \equiv \mathbf{ \mathrm{exp}: \int \hat{M}(\tau): }= \mathbb{I} + \sum^{\infty}_{n=1}  \int^{\tau}_{\tau_{0}} d \tau_{1} \int^{\tau_{1}}_{\tau_{0}} d \tau_{2}  \cdot \cdot \cdot \int^{\tau_{n-1}}_{\tau_{0}} d \tau_{n}  \hat{M}(\tau_{1}) \hat{M}(\tau_{2}) \cdot \cdot \cdot \hat{M}(\tau_{n}). \label{eq:wideeq}
\end{equation}
\end{widetext}
%\end{multicols}
%\begin{multicols}{2}

In scattering problems, $\mathbf{r}_{0}=\mathbf{r} (\tau_{0})$ is a vector with plane waves as components; $A(\tau)$ and $\alpha(\tau)$ are set by causal conditions of reflection and transmission. In the case of bound states $\mathbf{r}(\tau_{0}) \rightarrow \mathbf{0}$ when $\tau_{0} \rightarrow \pm \infty$, which is the case only for some values of $E$ substituted in $\hat{M}(\tau)$. The general solution can be simplified by performing the product of matrices \footnote{The sums resemble a Dyson series. The first term is defined as 1 for the diagonal entries of $\hat{R}$ and 0 for off-diagonal elements.}:

\begin{align}
\hat{R}(\tau)=\begin{pmatrix}
\sum\limits_{n \, \mathrm{even}} \int D\tau  \prod\limits^{n/2}_{k=1}  f_{2k-1}  & \sum\limits_{n \, \mathrm{odd}}   \int D\tau \prod\limits^{(n-1)/2}_{k=1} f_{2k-1} \\
\sum\limits_{n \, \mathrm{odd}}   \int D\tau \prod\limits^{(n-1)/2}_{k=1} f_{2k} & \sum\limits_{n \, \mathrm{even}}  \int D\tau \prod\limits^{n/2}_{k=1} f_{2k}
\end{pmatrix}
\end{align}
where
\begin{equation}
   \mathbf{r} =\hat{R} \, \mathbf{r}_0, \quad f_{n}=V(\tau_{n})-E
\end{equation}
and in Feynman's style we have
\begin{equation}
\int D\tau=\int^{\tau}_{\tau_{0}} d \tau_{n} \int^{\tau_{n}}_{\tau_{0}} d \tau_{n-1} \int^{\tau_{n-1}}_{\tau_{0}} d \tau_{n-2}  \cdot \cdot \cdot \int^{\tau_{2}}_{\tau_{0}} d \tau_{1}.
\end{equation}

The lower matrix elements of $\hat R$ in the asymptotic limit define the behavior of the wave; it is therefore important to name such functions and, in particular, the modulus of their joint contribution. We define the \textit{spectral function} $\mathcal{F}$ in full analogy with the special functions of the previous cases, as the limits

\begin{align}
\mathcal{F}(E)=\lim_{\tau \rightarrow \pm \infty} \left | \sum\limits_{n \, \mathrm{odd}}   \int D\tau \prod\limits^{(n-1)/2}_{k=1} f_{2k} \right | \nonumber \\
+ \lim_{\tau \rightarrow \pm \infty} \left | \sum\limits_{n \, \mathrm{even}}  \int D\tau \prod\limits^{n/2}_{k=1} f_{2k} \right |.
\label{EQ40}
\end{align}

It is important to note that the boundary conditions for bound states imply a vanishing $\mathcal{F}$. These conditions correspond to the evolution of the wave function along $x$. Meanwhile, the conditions in the upper row of $\hat R$, corresponding to the evolution of the derivative, are automatically satisfied due to the Pr\"ufer analysis criterion in Sobolev-type functions (in other words, they are redundant). The function \ref{EQ40} defines the eigenvalues of the system through the transcendental equation
\begin{equation}
\mathcal{F}(E_{n  })=0, \; n=0,1,..., n_{\mathrm{max}}.
\end{equation}
This expression involves all the possibilities of exactly solvable problems. The set of solutions are listed here as discrete, but it is well-known that some pathological potentials can exhibit mixed behavior \cite{vonneumann}, such as bound states in the continuum.
\subsection{Non linear case}
We address the problem of bound states. In the interstitial region, eq.(\ref{GE}) is 
\begin{equation}
\frac{d^{2} \phi (\tau)}{d \tau^{2}}=g \phi^{3}(\tau)-E\phi(\tau),
\label{EQ42}
\end{equation}
with $\phi (\tau)$ real and bounded, and $E<0$. The solution for this equation is $\phi(\tau)=A\, \mathrm{sn}(k \tau - \alpha;m)$. Similarly to the linear case, the boundary conditions are
\begin{equation}
A_{i+1} \mathrm{sn} (k_{i+1}\tau-\alpha_{i+1}; m_{i+1})=A_{i} \mathrm{sn} (k_{i}\tau-\alpha_{i}; m_{i}),
\end{equation}
\begin{align}
\nonumber
& A_{i+1} k_{i+1} \mathrm{cn} (k_{i+1}\tau-\alpha_{i+1}; m_{i+1}) \mathrm{dn} (k_{i+1}\tau-\alpha_{i+1}; m_{i+1})\\ \nonumber
& =A_{i} k_{i} \mathrm{cn} (k_{i}\tau-\alpha_{i}; m_{i}) \mathrm{dn} (k_{i}\tau-\alpha_{i}; m_{i}) \\
& +V_{i} \delta \tau A_{i} \mathrm{sn} (k_{i}\tau-\alpha_{i}; m_{i}). &
\end{align}
We apply the continuous limit again, with the aim of writing a closed differential equation, 
\begin{eqnarray}
\nonumber
& A(\tau+\Delta \tau) \mathrm{sn}[\mathrm{k(\tau+\Delta \tau)} \tau-\alpha(\tau+\Delta \tau); m(\tau+\Delta \tau)] \\
&=A(\tau) \mathrm{sn}[\mathrm{k(\tau)} \tau-\alpha(\tau); m(\tau)],
\label{EQ45}
\end{eqnarray}
\begin{align}
\nonumber
& A(\tau+\Delta \tau) k(\tau+\Delta \tau) \mathrm{cn}[\mathrm{k(\tau+\Delta \tau)} \tau-\alpha(\tau+\Delta \tau); m(\tau+\\ \nonumber
& \Delta \tau)] \mathrm{dn}[\mathrm{k(\tau+\Delta \tau)} \tau-\alpha(\tau+\Delta \tau) m(\tau+\Delta \tau)]=A(\tau) k(\tau) \\ \nonumber & \mathrm{cn}[\mathrm{k(\tau)} \tau-\alpha(\tau); m(\tau)] \mathrm{dn}[\mathrm{k(\tau)} \tau-\alpha(\tau) m(\tau)]+ V(\tau) \Delta \tau \\
& A(\tau) \mathrm{sn}[\mathrm{k(\tau)} \tau-\alpha(\tau); m(\tau)].&
\label{EQ46}
\end{align}
Following similar steps that led to (\ref{EQ34}) from  (\ref{EQ33}), the matching conditions (\ref{EQ45})  and (\ref{EQ46}) can be Taylor-expanded to first order in $\Delta \tau$ and thus establish an ordinary differential equation; we show the details in Appendix \ref{A}. Therefore, we obtain
\begin{equation}
    \frac{d \mathbf{r}(\tau)}{d \tau}=\hat{M_g}(\tau) \mathbf{r}(\tau),
    \label{EQ47bis}
\end{equation}
with
\begin{equation}
\hat{M_g} (\tau)=\begin{pmatrix}
0 & 1\\
V(\tau)-E+g \psi^{2}(\tau) & 0
\end{pmatrix}, \; \mathbf{r} (\tau)=\begin{pmatrix}
\psi(\tau)\\
\varphi (\tau)
\end{pmatrix}
\end{equation}
and
\begin{align}
& \psi(\tau)=A(\tau) \mathrm{sn} [k(\tau)\tau-\alpha(\tau);m(\tau)], \; \varphi (\tau)=A(\tau) \times \nonumber \\
& \times k(\tau) \mathrm{cn} [k(\tau)\tau-\alpha(\tau);m(\tau)] \mathrm{dn} [k(\tau)\tau-\alpha(\tau);m(\tau)].
\end{align}
In two-component form, this equation is re-written as a linear term and a source term 
\begin{align}
    \frac{d}{d \tau} \begin{pmatrix}
\psi(\tau)\\
\varphi(\tau)
\end{pmatrix}-\begin{pmatrix}
0 & 1\\
V(\tau)-E & 0
\end{pmatrix} \begin{pmatrix}
\psi(\tau)\\
\varphi(\tau)
\end{pmatrix} \nonumber \\
=g \psi^{2} \begin{pmatrix}
0 & 0\\
1 & 0
\end{pmatrix} \begin{pmatrix}
\psi(\tau)\\
\varphi(\tau)
\end{pmatrix}.
\label{EQ50}
\end{align}
We use the following ordered exponential that takes care of the linear part
\begin{eqnarray}
\hat{R_{0}}=\hat{R_{0}}(\tau)=\mathrm{exp} \, \left[: \int \begin{pmatrix}
0 & 1\\
V(\tau)-E & 0
\end{pmatrix} d\tau \, :\right], \\
\frac{d}{d \tau} \hat{R_{0}}^{-1}=-\hat{R_{0}}^{-1}  \begin{pmatrix}
0 & 1\\
V(\tau)-E & 0
\end{pmatrix}.
\end{eqnarray}
This is done in order to simplify the expressions in (\ref{EQ50}) such that 
\begin{eqnarray}
\hat{R_{0}} \frac{d}{d \tau} \left( \hat{R_{0}}^{-1} \begin{pmatrix}
\psi(\tau)\\
\varphi(\tau)
\end{pmatrix} \right)= \frac{d}{d \tau}  \begin{pmatrix}
\psi(\tau)\\
\varphi(\tau)
\end{pmatrix} \nonumber \\
-\begin{pmatrix}
0 & 1\\
V(\tau)-E & 0
\end{pmatrix} \begin{pmatrix}
\psi(\tau)\\
\varphi(\tau)
\end{pmatrix}.
\end{eqnarray}
Then our system resembles an \textit{interaction picture}, where the nonlinearity acts like the source or perturbation: 
\begin{eqnarray}
\hat{R_{0}} \frac{d}{d \tau} \left( \hat{R_{0}}^{-1} \begin{pmatrix}
\psi(\tau)\\
\varphi(\tau)
\end{pmatrix} \right)=g \psi^{2} \begin{pmatrix}
0 & 0\\
1 & 0
\end{pmatrix} \begin{pmatrix}
\psi(\tau)\\
\varphi(\tau)
\end{pmatrix}.
\end{eqnarray}
Left-multiplication by $\hat R_0$ and a redefinition of the spinor leads to the simplified system: 
\begin{align}
& \frac{d}{d \tau} \begin{pmatrix}
\tilde{\psi}\\
\tilde{\varphi}
\end{pmatrix}=g \psi^{2} \tilde{\sigma}_{-} \begin{pmatrix}
\tilde{\psi}\\
\tilde{\varphi}
\end{pmatrix},  \;
\begin{pmatrix}
\tilde{\psi}\\
\tilde{\varphi}
\end{pmatrix}= \hat{R_{0}}^{-1} \begin{pmatrix}
\psi\\
\varphi
\end{pmatrix}, \nonumber \\
& \tilde{\sigma}_{-}=\hat{R_{0}}^{-1} \begin{pmatrix}
0 & 0\\
1 & 0
\end{pmatrix} \hat{R_{0}}.
\end{align}
Both sides can be integrated in order to build an iterative integral series for a small time step $\Delta \tau$
\begin{align}
\begin{pmatrix}
\tilde{\psi}\\
\tilde{\varphi}
\end{pmatrix}_{\Delta \tau}-\begin{pmatrix}
\tilde{\psi}\\
\tilde{\varphi}
\end{pmatrix}_{0}= \int^{\Delta \tau}_{0} g \psi^{2} \tilde{\sigma}_{-} \begin{pmatrix}
\tilde{\psi}\\
\tilde{\varphi}
\end{pmatrix} d \tau \nonumber \\
= \Delta \tau g \psi^{2}(0) \tilde{\sigma}_{-}(0) \begin{pmatrix}
\tilde{\psi}(0)\\
\tilde{\varphi}(0)
\end{pmatrix} +O_{2} (\Delta \tau)\\
=\Delta \tau g \psi^{2}(0) \begin{pmatrix}
0 & 0\\
1 & 0
\end{pmatrix} \begin{pmatrix}
\psi(0)\\
\varphi(0)
\end{pmatrix} +O_{2} (\Delta \tau).
\label{eq56}
\end{align}
Without approximations we note that 
\begin{equation}
\mathrm{exp} \left( \Delta \tau g \psi^{2}(0) \begin{pmatrix}
0 & 0\\
1 & 0
\end{pmatrix} \right)=
\mathbb{I}+\Delta \tau g \psi^{2}(0) \begin{pmatrix}
0 & 0\\
1 & 0
\end{pmatrix},
\label{EQ57}
\end{equation}
which allows to write (\ref{eq56}) as 
\begin{eqnarray}
\begin{pmatrix}
\tilde{\psi}\\
\tilde{\varphi}
\end{pmatrix}_{\Delta \tau} \simeq \left( \mathbb{I}+\Delta \tau g \psi^{2}(0) \begin{pmatrix}
0 & 0\\
1 & 0
\end{pmatrix} \right) \begin{pmatrix}
\psi(0)\\
\varphi(0)
\end{pmatrix} \nonumber \\
\simeq \mathrm{exp} \left( \Delta \tau g \psi^{2}(0) \begin{pmatrix}
0 & 0\\
1 & 0
\end{pmatrix} \right) \begin{pmatrix}
\psi(0)\\
\varphi(0)
\end{pmatrix},
\end{eqnarray}
accurate to order 2, and our small-step solution reads 
\begin{eqnarray}
\begin{pmatrix}
\psi\\
\varphi
\end{pmatrix}_{\Delta \tau} \simeq \hat{R}_{0} \, \mathrm{exp} \left( \Delta \tau g \psi^{2}(0) \begin{pmatrix}
0 & 0\\
1 & 0
\end{pmatrix} \right) \begin{pmatrix}
\psi(0)\\
\varphi(0)
\end{pmatrix}.
\end{eqnarray}
It is advantageous to view this solution as a composition of two exponential maps corresponding to linear and nonlinear contributions. In general, the successive composition for infinitesimal time steps is the following ordered product
\begin{eqnarray}
\begin{pmatrix}
\psi_{n}\\
\varphi_{n}
\end{pmatrix}_{ \tau} \simeq \hat{R}_{0}(n,n-1)  \mathrm{exp} \left( \Delta \tau g \psi^{2}_{n-1} \begin{pmatrix}
0 & 0\\
1 & 0
\end{pmatrix} \right) \times  \nonumber \\ 
\hat{R}_{0}(n-1,n-2)  \mathrm{exp} \left( \Delta \tau g \psi^{2}_{n-2} \begin{pmatrix}
0 & 0\\
1 & 0
\end{pmatrix} \right) \times \cdot \cdot \cdot \nonumber \\
\cdot \cdot \cdot \hat{R}_{0}(1,0)  \mathrm{exp} \left( \Delta \tau g \psi^{2}_{0} \begin{pmatrix}
0 & 0\\
1 & 0
\end{pmatrix} \right)
\begin{pmatrix}
\psi(0)\\
\varphi(0)
\end{pmatrix}.
\end{eqnarray}
Finally, we apply the limit when $n \rightarrow \infty$, $\Delta \tau \rightarrow 0$ but $\tau$ finite. This leads to the familiar form of the \textit{Trotter limit} \cite{Wiebe_2010} \footnote{Sometimes also associated to Glauber, although the name Zassenhaus is mentioned more often when the Baker-Campbell-Hausdorff series temrinates.} for which the product of exponentials can be expressed as a single exponential map:
\begin{eqnarray}
     \hat{R}=\lim_{n \rightarrow \infty} \prod^{n}_{j=1} \hat{R}_{0}(j,j-1) e^{g \tau \sigma_{-} \psi^{2}_{n-1} / n} \\
   \equiv \exp{ \left[ :\int \mathbf{\hat M}(\tau): + g \tau \sigma_{-} \left( \sum^{n-1}_{j=0} \psi^{2}_{j} \right) \right] }.   
   \label{erre}
\end{eqnarray}
The instructions to utilize this formula for exact solutions to any desired order are iterative: First build $\hat R$ for a given $n$, say $n=1$, and apply it to the spinor $\Psi_1 = (\psi_1, \varphi_1)^{\rm{T}}$. Then, substitute $n=2$ in $\hat R$ and apply it to get $\Psi_2 = (\psi_2, \varphi_2)^{\rm{T}}$, and so on. The last function gives rise to a closed expression for $\hat R$ as in (\ref{erre}), whose elements are set as  
\begin{eqnarray}
\mathcal{F}_g(E) = \lim_{\tau \rightarrow \pm \infty} \left[ |\hat R_{21}|+|\hat R_{22}| \right],
\end{eqnarray}
and this is our spectral function for nonlinear problems $\mathcal{F}_g(E)=0$, whose roots yield the required solutions.

\section{Remarks on classical and quantum integrability}
It is clear that classical and quantum-mechanical equations of motion yield different results for general potentials beyond harmonic approximations.
Concretely, the quantum and classical cases differ in the solvability criteria: In one scenario, Newton's 2nd law is solved for canonical variables of the dynamical system and, in the other case, the wave equation is solved for a complex amplitude. However, the Heisenberg picture produces the same Hamilton equations, but now for operators; if there are not enough separation constants, neither classical nor quantum cases can be solved (e.g. by quadratures, but quantum-mechanically translates into a deficiency of conserved compatible operators). So even if the classical case allows solvability by quadratures --here the 1D system with conserved energy $H$ guarantees solutions by integrals-- we are left with a stationary wave equation that itself represents a non-trivial classical dynamical system, such as (\ref{EQ34}) and its nonlinear version (\ref{EQ47bis}) analyzed throughout the present paper. 

To show that this system is a nontrivial one, let us start with the stationary wave equation and build the dynamical system $x \rightarrow t$, $\psi (x) \rightarrow X(t)$, $\psi '(x) \rightarrow \dot{X}(t)=P(t)$ with the following nonautonomous form
\begin{equation}
\frac{d}{dt} \begin{pmatrix}
X\\
P
\end{pmatrix} = \begin{pmatrix}
0 & 1\\
E-V(t) & 0 
\end{pmatrix} \begin{pmatrix}
X\\
P
\end{pmatrix}.
\end{equation}
The energy $U=\frac{1}{2} (P^{2}+\left[E-V(t) \right] X^{2})$ is no longer conserved, so for arbitrary $V(t)$ this is, in general, a nonintegrable system in the classical sense, despite being a quantum system with conserved $H$\footnote{We highlight here that an effectively two-dimensional system without integrals of motion ($X$ and $t$ are two dimensions) does not predict the existence of solutions by separability, which makes the concept of quantum solvability and integrability more demanding.}. It can present nonlinear phenomena such as dynamical localization and chaos (also classical). The energy $U$ gives rise to the very famous Hill equation \cite{Magnus} when it is perturbed, and it is the subject of various stability studies when such small perturbation has the form $V(t)=V_{0}(t)+\delta V(t)$, based on specific behaviors of the corresponding solutions for $V_0(t)$ (see stability gaps or Arnold tongues \cite{Arnold, Broer}). There is a finite list of $V_{0}$ that are considered as solvable. Indeed, $\delta V$ can produce highly nontrivial effects, e.g. chaos, and requires the explicit evaluation of the spectral function $\mathcal{F}(E)$. So even though the GP equation has been formally solved for all possible potentials in this paper, the evaluation of $\mathcal{F}$ involves a high degree of complexity and some interesting surprises might await for us. 

Regards nonlinearities $V \rightarrow V+g|X|^{2}$, we have showed by construction that, indeed, it is possible to obtain a formal expression for waves and spectral functions that solve the problem. Two observations are in order:
\begin{itemize}
\item[a)] It is expected that every Liouville flow will solve any Hamiltonian dynamical system (formal solution) but that does not mean that the expression is easy to evaluate for energies $E$ that satisfy the boundary condition $X(t) \rightarrow X_{\infty}$, $t \rightarrow \pm \infty$. 

\item[b)] Quantum nonintegrability implies the classical one at the level of the equations followed by observables. However, quantum integrability seen as a complete set of compatible and conserved observables does not ensure analytically solvable wave equations. For this reason, there are some reservations when defining the concept of quantum chaos, and we work within more modest limits. A distinction proposed by Berry \cite{Berry}, "chaos" versus "chaology", deals with chaotic wavelike systems conceived as those whose classical limit yields a system that presents mixing, ergodicity and sensitivity to initial conditions.
\end{itemize}

A phase space diagram comparing the two cases $g\neq0$ and $g=0$ is provided in fig. \ref{FIG11}. The construction is explained in Appendix \ref{B}.

\begin{figure}[H]
 \includegraphics[width=\linewidth]{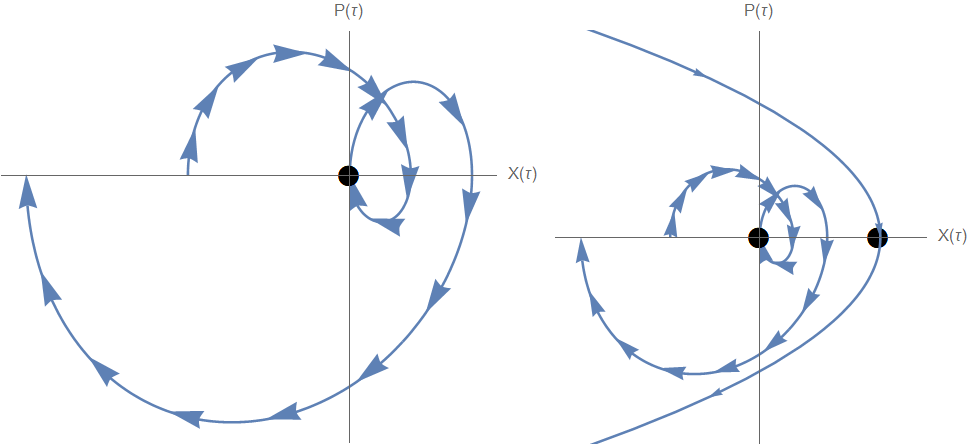}
    \centering
 \caption{Left panel: Linear stable equilibrium. Right panel: Nonlinear stable and unstable equilibria. When $\gamma<0$, the bright solitons (only bound states) correspond to a critical point $(X(\tau)=0,P(\tau)=0)$ in phase space. When $g>0$, we have dark solitons (unbounded). These correspond to points away from the origin, i.e., $\left( X(\tau)= \pm \sqrt{(E-V(\tau))/g},P(\tau)=0 \right)$.}
 \label{FIG11}
\end{figure}

\section{Conclusions}
In this work we have solved two paradigmatic problems of bound states for the nonlinear Schr\"odinger equation: a defect in the form of a Dirac delta (both negative and, surprisingly, positive with an attractive nonlinear interaction) and the square well potential. The equation that determines the allowed energies is solved using an improved graphical method for the GP equation, not reported in the literature. Then, we used the layered construction made of Dirac combs to show that the solution method for bound states is also effectivre for the nonlinear case. In this way, we approached the 1D equation with a general potential. In the continuous limit, a generalization of the Mello and Kumar equation\footnote{In the context of transport theory, this is known as Dorokhov-Mello-Pereyra-Kumar (DMPK) equation.} was found and we showed that the nonlinear form of these equations is identical to a mapping of the Gross-Pitaveskii equation to phase space. We also showed that any stationary Schr\"odinger equation, both linear and nonlinear, has formal solutions given by a series of ordered exponentials in the position variable; so technically any problem is solvable, not just the harmonic oscillator. 
We may conclude, from our discussions, that the complexity of the general problem resides in the spectral equations $\mathcal{F}(E)=0$, where $\mathcal{F}$ is expressed as a limit. To find the corresponding roots will sometimes require numerical techniques of evaluation, as the ordered series may not have a recognizable form. However, we note that, by construction, the wave functions are closed expressions for any potential and this settles the issue of exact solvability for 1D integrable quantum mechanical systems.

\section*{Acknowledgments}

Financial support from CONAHCYT under Grant No. CF-2023-G-763 is acknowledged. M.M. is grateful to VIEP-BUAP for support under Project No. 100518931-BUAP-CA-289.

\appendix{
\section{Deduction of equations for the nonlinear case \label{A}}
The Jacobi elliptic function $\phi_{i}(\tau)=A_{i} \mathrm{sn} (k_{i}\tau-\alpha_{i}; m_{i})$  satisfies the differential equation 
\begin{equation}
    \frac{d^{2} \phi_{i}(\tau)}{d \tau^{2}}=\frac{2 m^{2}_{i} k_{i}^{2}}{A_{i}^{2}} \phi_{i}^{3}(\tau)-k_{i}^{2} (1+m_{i}^{2}) \phi_{i}(\tau)
    \label{EQ66}
\end{equation}
If we compare the equations (\ref{EQ42}) and  (\ref{EQ66}), we obtain $\frac{2 m^{2}_{i} k^{2}_{i}}{A^{2}_{i}}=g$ and $E=k^{2}_{i} (1+m^{2}_{i})$, i.e., $\frac{2 m^{2}(\tau) k^{2}(\tau)}{A^{2}(\tau)}=g$ and $E=k^{2}(\tau) (1+m^{2}(\tau))$.

We expand the eqs. (\ref{EQ45}) and (\ref{EQ46}) to first order in $\Delta \tau$, and retain the derivatives, which give the following answers:
\begin{align}
& \frac{d}{d \tau} \left[ A(\tau)  \mathrm{sn} \left(k(\tau) \tau-\alpha(\tau); m(\tau)  \right) \right]=k(\tau) A(\tau) \mathrm{cn} \left(k(\tau) \tau \right. \nonumber \\
& \left.-\alpha(\tau); m(\tau)  \right) \mathrm{dn} \left(k(\tau) \tau-\alpha(\tau); m(\tau)  \right), \\
 &   \frac{d}{d \tau} \left[ k(\tau) A(\tau) \mathrm{cn} \left(k(\tau) \tau -\alpha(\tau); m(\tau)  \right) \mathrm{dn} \left(k(\tau) \tau \right.  \right. \nonumber \\
& \left. \left.-\alpha(\tau); m(\tau)  \right)  \right]=k^{2}(\tau) A(\tau) \mathrm{sn} \left(k(\tau) \tau-\alpha(\tau); m(\tau)  \right) \times \nonumber \\
& \left[ 2m^{2}(\tau) \mathrm{sn}^{2} \left(k(\tau) \tau-\alpha(\tau); m(\tau)  \right)-(1+m^{2}(\tau)) \right].
\end{align}
We define the wave functions as
\begin{equation}
    \psi(\tau)=A(\tau) \mathrm{sn} [k(\tau)\tau-\alpha(\tau);m(\tau)],
\end{equation}
 \begin{align}
  &   \varphi (\tau)=A(\tau) k(\tau) \mathrm{cn} [k(\tau)\tau-\alpha(\tau);  
  m(\tau)] \times \nonumber \\
&  \mathrm{dn} [k(\tau)\tau-\alpha(\tau);m(\tau)],
 \end{align}
 such that
\begin{align}
 &   \frac{d }{d \tau} \begin{pmatrix}
\psi(\tau)\\
\varphi (\tau)
\end{pmatrix}=  \\
& { \small \begin{pmatrix}
0 & 1\\
V(\tau)-k^{2}(\tau) (1+m^{2}(\tau))+\frac{2 m^{2}(\tau) k^{2}(\tau)}{A^{2}(\tau)} \psi^{2}(\tau) & 0
\end{pmatrix} } \begin{pmatrix}
\psi(\tau)\\
\varphi (\tau)
\end{pmatrix} \nonumber.
\end{align}
Then, we obtain the eq. (\ref{EQ65}). 

\section{Evolution in phase-space and critical points for the nonlinear problem \label{B}}
In the nonlinear case, we consider a first order nonlinear spinor differential equation.
\begin{equation}
    \frac{d }{d \tau} \begin{pmatrix}
\psi(\tau)\\
\varphi (\tau)
\end{pmatrix} =\begin{pmatrix}
0 & 1\\
V(\tau)-E+g \psi^{2}(\tau) & 0
\end{pmatrix} \begin{pmatrix}
\psi(\tau)\\
\varphi (\tau)
\end{pmatrix}.
\label{EQ65}
\end{equation}

If we want to analyze this system in phase space, we must define $X(\tau)=\psi (\tau)$ and $P(\tau)=\varphi (\tau)$. The resulting nonautonomous system evolves in time with the possibility of self-intersecting trajectories in $(X,P)$. The critical points are such that $d X(\tau)/d \tau=0$ and $d P(\tau)/d \tau=0$, and define the existence of bright and dark solitons. Then $P(\tau)=0$ and $(V(\tau)-E) X(\tau)+g X^{2}(\tau)=0$. For the linear case, the only critical point is $(X(\tau)=0,P(\tau)=0)$ and this corresponds to the bound state. In the general case, the critical point $(X(\tau)=0,P(\tau)=0)$ is the bright soliton with $g<0$; moreover, the points $\left( X(\tau)= \pm \sqrt{(E-V(\tau))/g},P(\tau)=0 \right)$ are the dark solitons with $g>0$, see fig. \ref{FIG11}.

}

%\section*{References}

%\end{multicols}

%\medline
%\begin{multicols}{2}
%%%%%%%%%%%%%%%
%
%Using BibTeX
\nocite{*}
\bibliography{ref}
%
%%%%%%%%
%
%Introducing references manually
%

%\begin{thebibliography}{99}
%\bibitem{ELK} E. Ley-Koo, Recent progress in confined atoms and molecules: Superintegrability and symmetry breakings, Rev. Mex. Fis. 64 (2018) 326, \url{https://doi.org/10.31349/RevMexFis.64.326}
%
%\bibitem{Griffiths} D.J. Griffiths, Introduction to Electrodynamics, 2nd ed. (Prentice Hall, Englewood Cliffs, NJ, 1989), pp. 331–334.
%
%\end{thebibliography}
%\end{multicols}

\end{document}